\documentclass[conference]{IEEEtran}
\IEEEoverridecommandlockouts
\usepackage{cite}
\usepackage{amsmath,amssymb,amsfonts}
\usepackage{algorithmic}
\usepackage{graphicx}
\usepackage{textcomp}
\usepackage{xcolor}
\def\BibTeX{{\rm B\kern-.05em{\sc i\kern-.025em b}\kern-.08em
    T\kern-.1667em\lower.7ex\hbox{E}\kern-.125emX}}
\usepackage{multirow}
\usepackage{hyperref}
\newcommand{\figdir}{./figures/}

\begin{document}

\title{Parallel Performance of ARM ThunderX2 for Atomistic Simulation Algorithms\\
}

\author{\IEEEauthorblockN{1\textsuperscript{st} William Robert Saunders}
\IEEEauthorblockA{\textit{Department of Physics} \\
\textit{University of Bath}\\
Bath, UK \\
wrs20@bath.ac.uk}
\and
\IEEEauthorblockN{2\textsuperscript{nd} James Grant}
\IEEEauthorblockA{\textit{Digital, Data and Technology} \\
\textit{University of Bath}\\
Bath, UK \\
rjg20@bath.ac.uk}
\and
\IEEEauthorblockN{3\textsuperscript{rd} Eike Hermann M\"uller}
\IEEEauthorblockA{\textit{Department of Mathematics} \\
\textit{University of Bath}\\
Bath, UK \\
em459@bath.ac.uk}}

\maketitle

\begin{abstract}
Atomistic simulation drives scientific advances in modern material science and accounts for a significant proportion of wall time on High Performance Computing facilities. 
It is important that algorithms are efficient and implementations are performant in a continuously diversifying hardware landscape. 
Furthermore, they have to be portable to make best use of the available computing resource. 

In this paper we assess the parallel performance of some key algorithms implemented in a performance portable framework developed by us.
We consider Molecular Dynamics with short range interactions, the Fast Multipole Method and Kinetic Monte Carlo.
To assess the performance of emerging architectures, we compare the Marvell ThunderX2 (ARM) architecture to traditional x86\_64 hardware made available through the Azure cloud computing service.
\end{abstract}


\section{Introduction}
High performance computing (HPC) hardware, which until recently has been dominated by Intel (CPU) and Nvidia (GPU) chipsets, is undergoing a rapid diversification.  
This is particularly evident in accelerators where Intel, AMD and Nvidia have been joined by competitors such as Graphcore and Google's TPU cloud service.
In the CPU space a resurgent AMD EPYC processor has made market share gains, while ARM is emerging as a genuine contender with the Marvell Thunder X2 processor which powers the UK Tier-2 HPC Isambard cluster (run by GW4 and the Met Office), and Amazon's Graviton2 chip, which is another ARM offering.
Particularly keenly anticipated has been Fujitsu's A64FX processor which powers the new Fugaku supercomputer that has taken the top spot in the Top500 list and 4$^\text{th}$ place in the Green500 list at the time of writing.

The arrival of Fugaku also marks a significant step towards exascale which is seeing significant research investment globally.
Programmes such as the Exascale Computing Project (US), the European High-Performance Computing Joint Undertaking (EU) and ExCALIBUR (UK) are exploring hardware and software approaches needed to attain and exploit exascale.
One particularly promising approach are domain specific languages which allow the automatic generation of performant code for a wide range of architectures.
The performance portable molecular dynamics \cite{Saunders2018} (PPMD) framework for molecular simulation, developed in our group, is an example of this.

To assess the capabilities of PPMD on a range of architectures, we present a performance comparison of three relevant algorithms and models implemented in PPMD.
We compare traditional HPC architectures based on Intel Skylake, AMD Naples/Rome with novel ARM hardware consisting of Marvell ThunderX2 CPUs.

\section{Background}\label{sec:background}
Computational modelling has become firmly established as a third approach to complement theory and experiments.
Over many years of research a range of algorithms have been developed by the scientific community to simulate processes that occur in materials.
These methods have been modified and extended to best leverage the available computing hardware.
In this section we introduce the three different algorithms studied in this paper and discuss the implementations that we used to compare the performance and parallel efficiency on four different chip architectures.

\subsection{Short-ranged Molecular Dynamics}\label{sec:lj}
The first method which we investigated is an important component of classical Molecular Dynamics (MD) calculations with short range interactions.
As a representative model system we consider liquid Argon, which can be modelled by a large number $N$ of particles contained in cuboid domain with periodic boundary conditions.
These particles interact through a pairwise potential, which is a function of the positions of the two interacting particles.
In MD the gradient of the potential determines the force between the two interacting particles.
At each step of a MD simulation the forces exerted on each atom by all other atoms are computed.
These forces are used to increment the momentum and update the position of the atoms by applying Newton's equations of motion.

For our numerical experiment we use the Lennard-Jones interatomic potential \cite{lj1}
\begin{equation}
    V_{LJ}(r) = 4\epsilon \left[\left( \frac{\sigma}{r} \right)^{12} - \left( \frac{\sigma}{r} \right)^6 \right],
\end{equation}
where $r$ is the interatomic distance between two interacting atoms.
The strength and shape of the potential are described by the constants $\epsilon$ and $\sigma$, which depend on the physics of the system under study.
For the rest of the paper we implicitly assume that all quantities are measured in suitable physical units.
Note that this potential and the corresponding force can be evaluated with 1 division and 25 additions/multiplications which is computationally cheap in comparison to alternative potentials.
For example, evaluating the Buckingham potential \cite{buck} involves square root and exponential operations.

In principle the computation of all $N(N-1)/2$ pairwise interactions is an $\mathcal{O}(N^2)$ operation.
To reduce the computational cost the interactions are typically approximated with a potential that is truncated at a given interatomic distance $r_c$.
In this case each particle only interacts with an on average fixed number of other particles which reduces the computational complexity to $\mathcal{O}(N)$.

An implementation of this MD algorithm must perform two main operations efficiently.
Firstly, all pairs of atoms that could be within $r_c$ of each other must be identified.
Secondly, for each of those pairs of atoms the interatomic potential must be evaluated. 
Since evaluation of the potentials is typically an operation with high arithmetic intensity, it will benefit from vectorisation on modern chip hardware.

A common technique to identify all interacting pairs of atoms is to decompose the simulation domain into cells with a side length that is at least $r_c$ and only consider interactions with atoms in neighbouring cells.
This domain decomposition (DD) allows parallelisation with a distributed memory approach.
We use the Message Passing Interface (MPI) to distribute the simulation domain over MPI ranks.
Since the neighbours of a particular atom are often stored on another MPI rank, each sub-domain is extended with a halo region.
For PPMD this is described in detail in \cite{Saunders2018a}.
Note that efficient algorithms exist for the nearest neighbour communication between adjacent sub-domains \cite{lammps1}.

Our implementation stores pairs of particles in a matrix based data structure, described by Rapaport \cite{rap11a}, known as a neighbour list.
This storage format is suitable for threaded shared memory environments such as CUDA and OpenMP.

The PPMD framework described in \cite{Saunders2018} exploits parallelism using MPI and OpenMP.
Pairwise- and single-particle- operations are described in a Domain Specific Language (DSL) implemented in Python.
Efficient C code is automatically generated for a range of target architectures.

\subsection{Long-ranged Molecular Dynamics}\label{sec:fmm}
In addition to short-ranged forces, MD simulations of charged atoms must also consider long-range electrostatic interactions.
In this case, the pairwise potential cannot be truncated at some fixed cutoff radius $r_c$.
A variety of different algorithms exist \cite{ewald1,under3, spme1, spme2, spme3} to compute these interactions with varying efficiency depending on the simulated system and computing hardware.

Here we use the Fast Multipole Method (FMM) \cite{3dfmm, 3dfmmz}, which is an $\mathcal{O}(N)$ algorithm, to compute long-ranged electrostatic interactions.
In the FMM algorithm a hierarchical grid is constructed on the simulation domain and the centre of each cell on all levels is the origin of a ``multipole'' expansion or ``local'' expansion.
The accuracy of the FMM depends on the number of terms used for these expansions.

At the start of the algorithm, multipole expansions of the charges in each of the cells on the finest level are constructed.
These multipole expansions are recursively translated and combined in the upward pass of the FMM.
In the downward pass multipole expansions are translated into local expansions which describe the potential field in the centre of the cell.
At the end of the tree traversal the local expansion in each cell describes the electrostatic potential induced by all charged atoms outside that cell and its nearest neighbours.
Finally, the electrostatic interactions between atoms in a cell and its nearest neighbours are computed by directly evaluating the pairwise Coulomb potential as discussed in the previous section.

An efficient implementation of the tree traversal must consider the translation and combination of expansions between cells in the grid hierarchy and manage distribution of work over CPU cores.
A more in-depth discussion of our FMM implementation is provided in \cite{Saunders2018a}.

\subsection{kinetic Monte Carlo}\label{sec:kmc}
Kinetic Monte Carlo (KMC) can be used to investigate time dependent phenomena such charge transport in semi-conductors.
In a rejection free KMC \cite{Thompson2018, Morgan2017, Groves2016} simulation of a photovoltaic cell or battery, $N$ charged atoms can hop between adjacent sites of a (generally unstructured) lattice.
At each step hops are considered for all mobile charges in the system but only one hop is accepted with a probability that is proportional to the so called ``hopping rate'' or ``propensity''.
In materials such as semi-conductors, the hopping rates are a function of the change in electrostatic energy of the moving charge.

In \cite{Saunders2019} we developed a new algorithm to perform the computation of these differences in electrostatic energy with $\mathcal{O}(N)$ computational complexity by using the local expansions described in the FMM algorithm.
KMC, along with our electrostatic algorithm, is very well suited to parallel architectures: Firstly, differences in electrostatic energy for all potential moves can be computed independently.
Secondly, when a move is accepted all local expansions can be updated to reflect the accepted move in parallel.
Only the final accepted move as to be communicated between all processors.

\section{Setup and Results}

\subsection{Platforms and Configuration}

We used two HPC facilities to assess the performance of PPMD.
An Azure based platform of traditional x86\_64 hardware and the UK Tier 2 Isambard facility.
This allows comparison between traditional x86 architectures and novel ARM processors while using state-of-the-art hardware.
In Table \ref{tab:platforms} the node configurations on both machines are provided.

\begin{table*}[htbp]
    \caption{Overview of node architectures. Peak floating point rates (Rpeak) are computed assuming the listed clock speed and FLOPs/cycle. Numerical experiments were performed using the listed number of MPI ranks and OpenMP threads for each node.}
    \label{tab:platforms}
    \begin{center}
        \begin{tabular}{|cc|c|c|c|c|c|c|c|c|}
            \hline
             & \textbf{Platform} & \textbf{Processor} & \textbf{Core}  & \textbf{Clock}        & \textbf{FLOPs/}&  \textbf{Bandwidth}& \textbf{Rpeak}           &\textbf{MPI rank} &\textbf{OpenMP thread}\\
             &                   &                    & \textbf{Count} & \textbf{Speed (GHz)}  & \textbf{cycle} &   \textbf{(GB/s)}  & \textbf{(Tflops)}  &\textbf{count}    &\textbf{count}        \\
            \hline
                                                                                        & Isambard  & Marvell ThunderX2     & 64     & 2.1  & 8  & 318   & 1.08 & 8   & 8  \\
            \multirow{3}{*}{Azure} \multirow{3}{*}{{\fontsize{50}{0}\selectfont \{}}    & HC44      & Intel Platinum 8168   & 44     & 1.9  & 32 & 191   & 2.68 & 4   & 11 \\
                                                                                        & HB60      & AMD EPYC 7551         & 60     & 2.0  & 8  & 263   & 0.96 & 15  & 4  \\
                                                                                        & HB120     & AMD EPYC 7V12         & 120    & 2.45 & 16 & 350   & 4.70 & 30  & 4  \\
            \hline
        \end{tabular}
    \end{center}
\end{table*}

On the Azure platform Intel MPI 2018.4 and the Intel Compiler 19.0.5 were used.
As Isambard system is a Cray XC50 system we used Cray mpich 7.7 along with the GNU compiler version 9.2.0.
The Cray compiler was not used for the Cray compiler cannot be called on compute nodes which is essential for code generation frameworks such as PPMD.
All codes were run as a hybrid MPI+OpenMP setup using the configurations listed in Table \ref{tab:platforms}.
We now compare the performance of the three benchmarks described in Section \ref{sec:background} on the more traditional chip architectures and the novel ARM ThunderX2 processor.

\subsection{Strong Scaling Comparison}

\begin{figure}[htbp]
    \centerline{\includegraphics{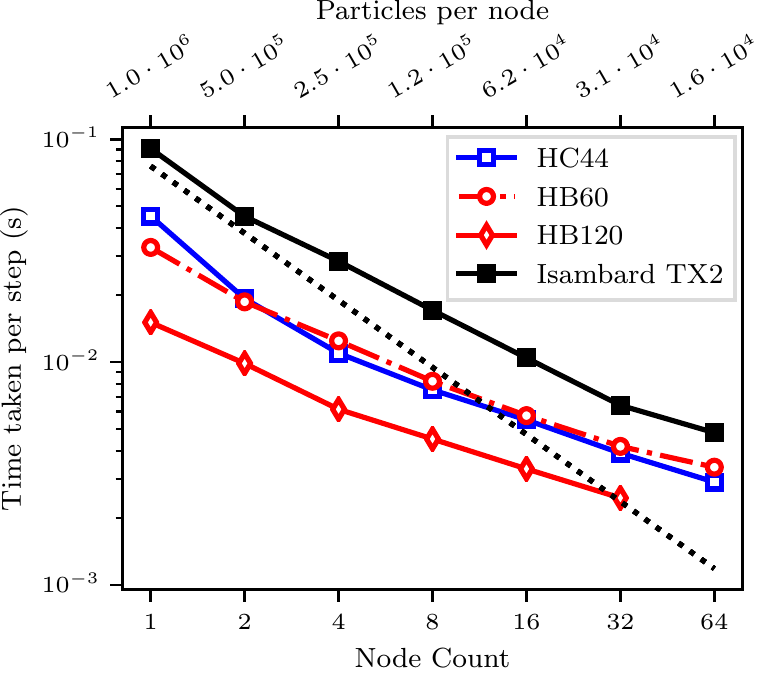}}
    \caption{Lennard-Jones strong scaling experiment. Time per step plotted against number of nodes for each HPC platform. Dashed black indicates ideal scaling.}
\label{fig:lj}
\end{figure}

For the Lennard-Jones benchmark (Section \ref{sec:lj}), $N=10^6$ atoms were arranged in a cubic lattice with spacing $a=0.945$.
The potential was truncated at a cutoff radius $r_c=2.5$ and we set $\sigma=1, \epsilon=1$.
The neighbour list was rebuilt at least every 10 steps with a $2.75$ cutoff.
In Figure \ref{fig:lj} we plot the time taken per simulation step, for each of our test platforms, as the number of nodes is increased while keeping $N$ fixed in a strong scaling experiment.

This particular benchmark is tailored to reach a strong-scaling limit at a relatively small node count.
In this limit the number of atoms per CPU core is relatively low $<$1000 hence, at this density and cutoff, the cost of potential evaluations is small in comparison to bookkeeping and communication.
In particular, this benchmark computes and communicates (\verb+MPI_Allreduce+) the potential and kinetic energy at each step.

Inspection of binaries reveals that on the Intel and AMD processors the Intel compiler produced AVX2 instructions.
On Isambard, GCC did not vectorise the loops that evaluate the potential.
To investigate further, we modified the generated code that evaluates the pairwise potential to include OpenMP SIMD pragmas and recompiled using the ARM compiler version 20.1.
SIMD instructions were emitted and the single node time per step improved to 0.081s from 0.091s.

\begin{figure}[htbp]
    \centerline{\includegraphics{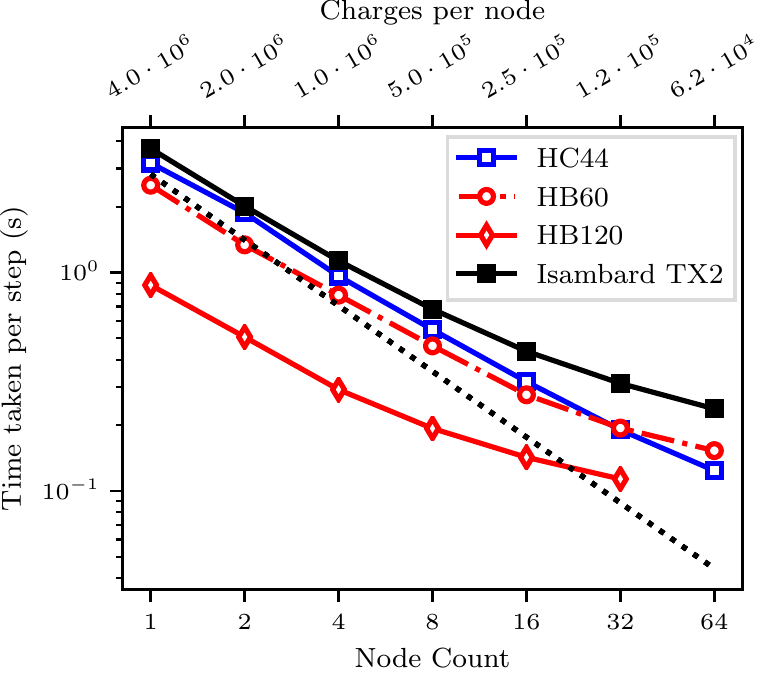}}
    \caption{FMM strong scaling experiment. Time per step plotted against number of nodes for each HPC platform. Dashed black indicates ideal scaling.}
\label{fig:fmm}
\end{figure}

For the FMM benchmark (Section \ref{sec:fmm}), $N=4 \cdot 10^6$ atoms were arranged in a cubic lattice with spacing $a=6.6$.
To prevent oppositely charged atoms from collapsing onto each other a repulsive short-range Lennard-Jones potential with cutoff $r_c = 4.0$ was added.
We configured the FMM with ten expansion terms and six levels in the hierarchical grid.
In Figure \ref{fig:fmm} we plot the time taken per simulation step, for each of our test platforms, as the number of nodes is increased while keeping $N$ fixed in a strong scaling experiment.

We observe that for the FMM benchmark GCC produced vector instructions for both the direct interactions and for the expensive multipole to local expansion operations.
In our FMM implementation the number of processors that can do useful work on each grid level in the upward and downward pass is limited by the total number of cells on that level.
If the number of cores is greater than the number of cells on a level then some cores will be idle, which impacts parallel scaling.

\begin{figure}[htbp]
    \centerline{\includegraphics{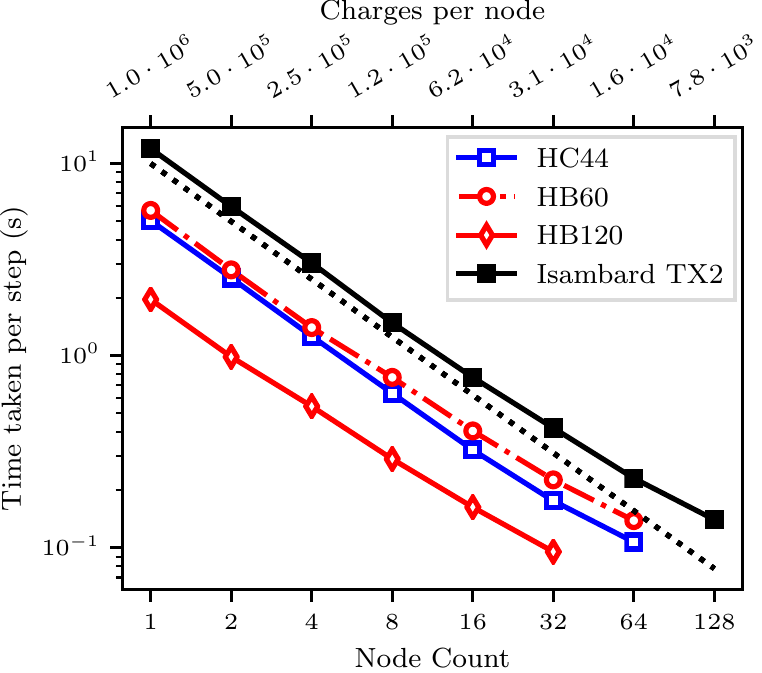}}
    \caption{KMC strong scaling experiment. Time per step plotted against number of nodes for each HPC platform. Dashed black indicates ideal scaling.}
\label{fig:kmc}
\end{figure}

Finally, for the KMC benchmark (Section \ref{sec:kmc}), $N=10^6$ charged atoms were placed on a cubic lattice.
At each lattice site atoms may hop in one of six directions assuming that the destination site is not occupied.
Consequently, at each step approximately $7 \cdot 10^6$ energy evaluations are performed.
The electrostatic solver was configured to use 12 expansion terms which, as is described in detail in \cite{Saunders2019}, is sufficient to achieve a mean relative error in the potential energy of $\approx 10^{-5}$.
In Figure \ref{fig:kmc} we plot the time taken per step for each of our test platforms as the number of nodes is increased while keeping $N$ fixed in a strong scaling experiment.

As expected, since the algorithm can be parallelised over particles, the KMC benchmark scales very well as the ratio of computation to communication is high.
The only communicated data are the details for the accepted move.
Even for the smallest sub-domains (at the largest node counts) there is enough computational work for the communication time to be negligible.
Inspection of binaries reveals that GCC did vectorise some expensive loops, such as bookkeeping operations.
However not all performance critical loops were vectorised.
We recompiled certain expensive operations with the ARM compiler, which produced vector instructions, but we did not observe a performance improvement.

\section{Conclusion}
In this paper we presented strong scaling results for the implementation of three algorithms highly relevant in atomistic simulation.
All algorithms were implemented in our performance portable PPMD framework.
In terms of strong scaling efficiency, the Isambard platform performed competitively in comparison to the Azure platform.
In our FMM benchmark, we observed that in terms of time-to-solution the ThunderX2 architectures was competitive with the x86 based chipsets which were using a very established compiler.
The results presented here show that further development of our code generation system (in collaboration with the compiler developers) is necessary to implement algorithms which are more efficient on the ARM hardware.
Future work should investigate, from a microarchitecture standpoint, the performance difference between the HB60 cores and the ThunderX2 cores which theoretically both perform 8FLOPs/cycle.
\section*{Acknowledgments}
This project has received funding from the European Union’s Horizon 2020 research and innovation programme under grant agreements No 646176 and No 824158. This research made use of the Balena High Performance Computing (HPC) Service at the University of Bath and the Isambard UK National Tier-2 HPC Service (\url{http://gw4.ac.uk/isambard/}). Isambard is operated by GW4 and the UK Met Office, and funded by EPSRC (EP/P020224/1). 
We thank the University of Bath ARC team and Red Oak Consulting for providing their expertise when running on the Azure platform.

\bibliographystyle{IEEEtran}
\bibliography{paper}

\begin{thebibliography}{10}
\providecommand{\url}[1]{#1}
\csname url@samestyle\endcsname
\providecommand{\newblock}{\relax}
\providecommand{\bibinfo}[2]{#2}
\providecommand{\BIBentrySTDinterwordspacing}{\spaceskip=0pt\relax}
\providecommand{\BIBentryALTinterwordstretchfactor}{4}
\providecommand{\BIBentryALTinterwordspacing}{\spaceskip=\fontdimen2\font plus
\BIBentryALTinterwordstretchfactor\fontdimen3\font minus
  \fontdimen4\font\relax}
\providecommand{\BIBforeignlanguage}[2]{{%
\expandafter\ifx\csname l@#1\endcsname\relax
\typeout{** WARNING: IEEEtran.bst: No hyphenation pattern has been}%
\typeout{** loaded for the language `#1'. Using the pattern for}%
\typeout{** the default language instead.}%
\else
\language=\csname l@#1\endcsname
\fi
#2}}
\providecommand{\BIBdecl}{\relax}
\BIBdecl

\bibitem{Saunders2018}
W.~R. Saunders, J.~Grant, and E.~H. M{\"u}ller, ``{A domain specific language
  for performance portable molecular dynamics algorithms},'' \emph{Computer
  Physics Communications}, vol. 224, pp. 119--135, 2018.

\bibitem{lj1}
J.~E. Lennard-Jones, ``Cohesion,'' \emph{Proceedings of the Physical Society},
  vol.~43, no.~5, p. 461, 1931.

\bibitem{buck}
\BIBentryALTinterwordspacing
R.~A. Buckingham and J.~E. Lennard-Jones, ``The classical equation of state of
  gaseous helium, neon and argon,'' \emph{Proceedings of the Royal Society of
  London. Series A. Mathematical and Physical Sciences}, vol. 168, no. 933, pp.
  264--283, 1938. [Online]. Available:
  \url{https://royalsocietypublishing.org/doi/abs/10.1098/rspa.1938.0173}
\BIBentrySTDinterwordspacing

\bibitem{Saunders2018a}
W.~R. Saunders, ``{Development of A Performance-Portable Framework For
  Atomistic Simulations},'' Ph.D. dissertation, University of Bath, 2018.

\bibitem{lammps1}
\BIBentryALTinterwordspacing
S.~Plimpton, ``{Fast Parallel Algorithms for Short-Range Molecular Dynamics},''
  \emph{Journal of Computational Physics}, vol. 117, no.~1, pp. 1 -- 19, 1995.
  [Online]. Available:
  \url{http://www.sciencedirect.com/science/article/pii/S002199918571039X}
\BIBentrySTDinterwordspacing

\bibitem{rap11a}
D.C.Rapaport, ``Enhanced molecular dynamics performance with a programmable
  graphics processor,'' \emph{Computer Physics Communications}, vol. 182, pp.
  926--934, 2011.

\bibitem{ewald1}
P.~P. Ewald, ``{Die Berechnung optischer und elektrostatischer
  Gitterpotentiale},'' \emph{Annalen der Physik}, vol. 369, no.~3, pp.
  253--287, 1921.

\bibitem{under3}
D.~Frenkel and B.~Smit, in \emph{{Understanding Molecular Simulation (Second
  Edition) }}.\hskip 1em plus 0.5em minus 0.4em\relax Academic Press, 2002, p.
  291.

\bibitem{spme1}
M.~Deserno and C.~Holm, ``{How to mesh up Ewald sums. I. A theoretical and
  numerical comparison of various particle mesh routines},'' \emph{The Journal
  of Chemical Physics}, vol. 109, no.~18, pp. 7678--7693, 1998.

\bibitem{spme2}
T.~Darden, D.~York, and L.~Pedersen, ``{Particle mesh Ewald: An $N\cdot
  \log(N)$ method for Ewald sums in large systems},'' \emph{The Journal of
  Chemical Physics}, vol.~98, no.~12, pp. 10\,089--10\,092, 1993.

\bibitem{spme3}
U.~Essmann, L.~Perera, M.~L. Berkowitz, T.~Darden, H.~Lee, and L.~G. Pedersen,
  ``{A smooth particle mesh Ewald method},'' \emph{The Journal of Chemical
  Physics}, vol. 103, no.~19, pp. 8577--8593, 1995.

\bibitem{3dfmm}
L.~Greengard, ``The rapid evaluation of potential fields in particle systems,''
  Ph.D. dissertation, Yale University, 1987.

\bibitem{3dfmmz}
L.~Greengard and V.~Rokhlin, ``A new version of the fast multipole method for
  the laplace equation in three dimensions,'' \emph{Acta Numerica}, vol.~6, pp.
  229 -- 269, 1997.

\bibitem{Thompson2018}
I.~R. Thompson, M.~K. Coe, A.~B. Walker, M.~Ricci, O.~M. Roscioni, and
  C.~Zannoni, ``{Microscopic origins of charge transport in triphenylene
  systems},'' \emph{Physical Review Materials}, vol.~2, no.~6, p. 064601, 2018.

\bibitem{Morgan2017}
B.~J. Morgan, ``Lattice-geometry effects in garnet solid electrolytes: {A}
  lattice-gas {Monte} {Carlo} simulation study,'' \emph{Royal Society Open
  Science}, vol.~4, no.~11, p. 170824, Nov. 2017.

\bibitem{Groves2016}
C.~Groves, ``{Simulating charge transport in organic semiconductors and
  devices: {A} review},'' \emph{Reports on Progress in Physics}, vol.~80,
  no.~2, p. 026502, 2016.

\bibitem{Saunders2019}
W.~R. Saunders, J.~Grant, E.~H. M{\"u}ller, and I.~Thompson, ``{Fast
  electrostatic solvers for kinetic Monte Carlo simulations},'' \emph{Journal
  of Computational Physics}, p. 109379, 2020.

\end{thebibliography}

\vspace{12pt}
\end{document}